Toward an Evolutionary-Predictive Foundation for Creativity

Commentary on "Human creativity, evolutionary algorithms, and predictive representations: The mechanics of thought trials" by Arne Dietrich & Hilde Haider, 2015

Liane Gabora
University of British Columbia

and

Stuart Kauffman
Institute for Systems Biology, Seattle

Correspondence regarding this article should be addressed to:

Liane Gabora
Department of Psychology, University of British Columbia
Okanagan Campus, 1147 Research Road, Kelowna BC, Canada V1V 1V7
Email: liane.gabora@ubc.ca

Abstract

Dietrich and Haider (2014) justify their integrative framework for creativity founded on evolutionary theory and prediction research on the grounds that "theories and approaches guiding empirical research on creativity have not been supported by the neuroimaging evidence". Although this justification is controversial, the general direction holds promise. This commentary clarifies points of disagreement and unresolved issues, and addresses mis-applications of evolutionary theory that lead the authors to adopt a Darwinian (versus Lamarckian) approach. To say that creativity is Darwinian is not to say that it consists of variation plus selection—in the everyday sense of the term—as the authors imply; it is to say that evolution is occurring because selection is affecting the distribution of randomly generated heritable variation across generations. In creative thought the distribution of variants is not key, i.e., one is not inclined toward idea A because 60% of one's candidate ideas are variants of A while only 40% are variants of B; one is inclined toward whichever seems best. The authors concede that creative variation is partly directed; however, the greater the extent to which variants are *generated* non-randomly, the greater the extent to which the distribution of variants can reflect not selection but the initial generation bias. Since each thought in a creative process can alter the selective criteria against which the next is evaluated, there is no demarcation into generations as assumed in a Darwinian model. We address the authors' claim that reduced variability and individuality are more characteristic of Lamarckism than Darwinian evolution, and note that a Lamarckian approach to creativity has addressed the challenge of modeling the emergent features associated with insight.

As Dietrich and Haider (2014) suggest, synthesizing creativity into an integrated framework that incorporates evolutionary theory and literature on prediction could enrich our understanding of how this mental faculty has transformed human civilizations. This commentary briefly addresses their negative pronouncement on efforts toward a neuroscience of creativity, clears out some errors in their application of evolutionary theory to creativity, and clarifies points of disagreement and unresolved issues in order to pave the way for an integrated understanding of how the creative process works.

**Prospects for a Neuroscience of Creativity**

The rationale Dietrich and Haider give for embarking on the research program described in this paper is that "all theories and approaches guiding empirical research on creativity—divergent thinking, defocused attention, right brains, low arousal, prefrontal activation, alpha enhancement, etc.—have not been supported by the neuroimaging evidence." We believe it is important to point out that two recent quantitative meta-analyses of neuroimaging data show remarkable process specificity in the neural correlates of creativity. The first showed that the neural correlates analogy and metaphor—both of which are related to creativity—exhibit consistent and dissociable patterns of activation across fMRI studies (Vartanian, 2012). The second showed that the neural correlates of creative cognition are consistently dissociable as a function of the task (i.e., generation vs. combination) or modality (i.e., verbal vs. nonverbal) across studies (Gonen-Yaacovi, de Souza, Levy, Urbanski, Josse, & Volle, 2013). It has been suggested that negative pronouncements on the field stem from the manner in which reviews have been carried out. Kounios and Beeman (2014: p. 73) write:

> [O]ne recent review of cognitive neuroscience research on creativity and insight lumps together widely diverse studies characterized by a variety of definitions, assumptions, experimental paradigms, empirical phenomena, analytical methods, and stages of the solving process (and inconsistent experimental rigor). Unsurprisingly, because of such indiscriminate agglomeration, that review failed to find much consistency across studies, leading the authors to pronounce a negative verdict on the field (Dietrich & Kanso 2010).

Interestingly, both those who believe the neuroscience of creativity is progressing and those such as Dietrich and Haider who "consider the current paradigm as failed" agree that the concepts and component processes by which we characterize creativity are in need of refinement. Thus although the rationale Dietrich and Haider give for their project is based on a conclusion that is controversial, their proposed new direction for creativity research may be meritorious.

**The Applicability of a Darwinian framework**

Dietrich and Haider continue in a direction initiated by Campbell's (1960) blind variation and selective retention (BVSR) theory in which creativity is modeled as a Darwinian process. They deviate from Campbell in claiming that a Darwinian framework is appropriate because creativity entails partial coupling of variation to selection. To examine this claim we must examine the issue of when a Darwinian framework is appropriate. The domain of applicability of natural selection can be appreciated when one considers that Darwin was faced with the paradox of how species *accumulate* change when traits acquired over organisms' lifetimes are *obliterated.* For example, parents do not pass on acquired traits such as tattoos or knowledge of chess. Darwin insightfully shifted the focus from the organism to the population, noting that there is population-level change over generations in, not acquired traits, but inherited traits. Although acquired traits are discarded, inherited traits are not. When random mutations to inherited traits are beneficial for their bearers, their bearers have more offspring, and these traits proliferate at the expense of inferior ones. Thus species accumulate change despite that traits acquired over their lifetimes are obliterated because over generations selection can lead to substantial change in the distribution of heritable traits across the population. Since in the present context the entities are not organisms but ideas, the traits in question are characteristics of candidate ideas one is entertaining, and the "offspring" of a given idea are presumably more refined versions one might have following initial conception of the idea. Note that (contrary to what is claimed in the article) it is not necessary that "individual units of a population are unique". There could be millions of identical individuals of one type and millions of identical individuals of another type; all that is necessary is that there be two or more types.

Darwin himself claimed that his theory was not a theory about how new forms *come into being*; but a theory about the effect of selection on the *distribution* of variants over time. According to standard neoDarwinism as formalized by Fisher (1930), Wright (1931), and Haldane (1932), to say that creativity is Darwinian is not to say that it consists of variation plus selection as the authors imply; it is to say that evolution is occurring because *selection is affecting the distribution of variants across generations*. An example of selection as the term is used in everyday, nontechnical parlance is: you have three ideas and choose the best. An example of Darwinian selection is:

(1) You generate 30 candidate ideas: 10 each of type A, B, and C.
(2) You expose them to selection criteria (tests of their effectiveness), which affect how many "offspring ideas" they bear.
(3) In the next generation there are 15 ideas of type A, 5 of type B, and 10 of type C.
(4) After several generations all your candidate ideas are variants of type A.

We know of no data that suggests that in creative thought the distribution of variants is the key issue. One is not inclined toward idea A because 60% of one's candidate ideas are variants of A while only 40% are variants of idea B; one is inclined toward whichever of A or B seems best. Thus we believe it is important to clarify that while selection may play a role in creativity, it is selection in the everyday, nontechnical sense as opposed to Darwinian selection.

**Acquired versus inherited traits**

Note that at the core of Darwin's theory is a distinction between acquired traits that are not passed on and inherited traits that are, for selection acts on heritable differences. The effect of transmission of acquired traits can be seen by comparing Figure 1a—a schematic depiction of natural selection—to Figure 1b—in which traits acquired by variants in one generation are passed down vertically to the next. Because thousands or millions of traits can be acquired *within* a generation, if they *are* transmitted, this can easily wipe out the effect of natural selection, which occurs *between* generations.

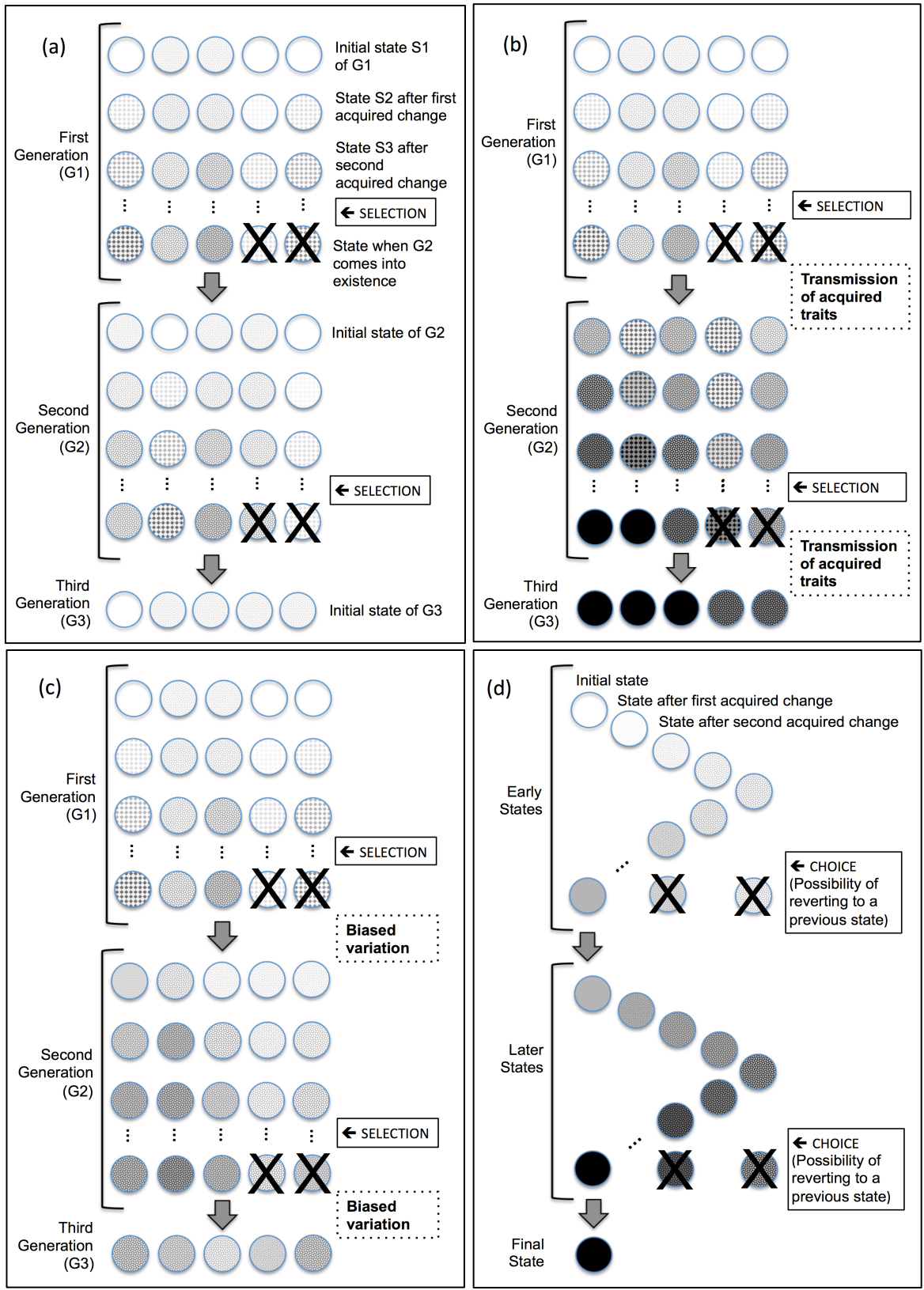

Caption for Figure 1. (a) Evolution through natural selection. (b) Evolution with transmission of acquired traits. (c) Evolution with biased variation (e.g., 'directed mutation'). (d) Sequential evolution of an entity with each change of state potentially altering the fitness criteria that guide the next change of state. The significance of the different rows in (b) and (c) is identical to that in (a) but the explanations have been removed in order to make the figure less cluttered. Three dots are used to represent that many more changes may be acquired than can be indicated in such a figure.

If traits are *both* acquired and inherited—as one might be tempted to argue is the case with respect to creative change—the effect of natural selection is most likely negligible because acquired change (which can take place in an instant) can accumulate exponentially faster than inherited change (which requires generations). If acquired change is not discarded it overwhelms Darwin's population-level mechanism of change; it 'swamps the phylogenetic signal' (Tëmkin, & Eldredge, 2007). It is only because, in biological evolution, changes due to the faster process (the effects of learning, aging, and so forth) are in general not retained from one generation to the next that changes due to the slower process (the effects of mutation and recombination) are not drowned out, and play an evolutionary role. In biology it is increasingly realized that the relationship between inherited and acquired traits is more complex than Darwin could have anticipated[1], and accordingly Darwin's theory of natural selection is limited even as a model of organic evolution (Kauffman, 1993; Koonin, 2009; Koonin & Wolf, 2012; Woese, 2002; Woese & Goldenfeld, 2009). This suggests that caution is needed in applying the theory beyond the domain for which it was intended.

Note that this discussion concerns the conditions that render a particular theory—specifically, the theory of natural selection—appropriate as a model of the phenomenon at hand, e.g., a set of empirical facts about the biological or cultural world. Any model is a simplification. Sometimes the simplifying assumptions of the model capture the gist of the phenomenon, and it is useful. Other times the phenomenon deviates significantly from the simplifying assumptions of the model, and the model is misleading. To the extent that acquired traits are transmitted the theory of natural selection can lead us astray.

**Biased Variation / Directed Mutation**

As the authors point out, one can sometimes anticipate the criteria that will be used to evaluate creative ideas, and use them to bias or "direct" the generation of ideational variants away from random (see also Eysenck, 1993). However, according to standard neoDarwinian theory as developed by Fisher (1930), Wright (1931), and Haldane (1932), a central assumption of natural selection is that however variation arises it is not correlated with its prospective selective advantage. The greater the extent to which variants are *generated* non-randomly, the greater the extent to which it is possible that the distribution of variants reflects *not* selection (which occurs *after* variants have come into existence) but the *initial generation bias*. Therefore, the less applicable natural selection may be as a model. This is illustrated in Figure 1c. The argument is analogous to the one in the previous section concerning transmission of acquired traits; if variants are generated using such factors as strategy or intuition this can render negligible any effect of selection on the distribution of variants across generations.

---

[1] For example, epigenetic markings of DNA by methylation can alter gene expression trans-generationally, enabling organisms with adaptive epigenetic alterations in gene expression to select for mutations that lock in these beneficial alterations (Nakayama, Rice, Strahl, Allis, & Grewal, 2001).

The effect of nonrandom variation on the applicability of a Darwinian framework applies to any evolutionary process, biological or cultural. The assumption of randomness in biology generally holds well enough to serve as a useful approximation, but in fact biological variation is not genuinely random; for example, a bias might arise due to mutagenic agents, assortative mating, or in the case of synthetic biology, directed mutation (e.g., Currin, Swainston, Day, Kell, 2014, in press). Natural selection works by generating *numerous* possibilities through a process that can be approximated by a random distribution, such that often enough at least one variant is bound by chance to be fitter than what came before, and likely to be amplified in subsequent generations.[2] Human creativity works by generating *few* possibilities using intelligence, foresight, and intuition, such that they are more likely than chance to be adaptive. Creativity is not in need of a theory that was developed to describe situations involving large numbers of possibilities generated through processes that can be approximated by a random distribution. Thus, to the extent that variation in human creative outputs is biased from a random distribution, change over generations is not necessarily due to competitive exclusion of inferior variants; in other words, it is not due to the mechanism offered by Darwin. Indeed if this initial biasing correctly anticipates the actual selection pressures—i.e., if there is 1:1 coupling of variation to selection—the variation phase has already done the work, so evolution is not due to selection.

Thus, while Dietrich and Haider claim that criticism of Darwinian approaches to creativity "only holds… if we insist on total blindness" (p. 901), the criticisms are actually subtler. The argument is not merely that a Darwinian theory of creativity is inappropriate because creativity is not completely blind. The argument is: to the extent that creativity is not blind, a Darwinian framework is inapplicable.

**Sequential Selection**

Another factor that limits the applicability of a Darwinian framework is that each thought or idea can modify the selection criteria by which the next thought or idea is evaluated (Gabora, 2011a,b), as illustrated in Figure 1d. One might occasionally entertain multiple possibilities at once (as when an artist generates a collection of sketches and then chooses amongst them). However, in generating any particular sketch the artist was free to draw upon observations made during previous sketches. Therefore, they are not evaluated according to the same selection criteria, and there is no basis upon which to claim that one generation is ending and the next is beginning. Moreover, the artist could reconsider ideas that arose arbitrarily far back in time—i.e., revive ideas that were seemingly "dead"—but because Darwinian selection is predicated on the notion of generations (Fisher, 1930) it is incompatible with this kind of "bringing the dead back to life" (thus, to avoid confusion, in Figure 1d the term 'choice' is used rather than 'selection').

---

2 Note that the issue of "often enough" is the vexing problem of the "arrival of the fitter". In biological evolution, arrival of the fitter often entails the finding of an opportunistic new use for an existing structure or process, referred to as preadaptation. However, even in preadaptation there is no way to pre-state what new functionality will be selected at the level of the whole organism in its world, so the new use can be treated as random.

**Examining the argument that creativity is not Lamarckian**

Lamarckism is commonly understood to refer to an evolutionary process in which traits acquired during an organism's lifetime can be transmitted to offspring without the mechanism of genetics (Mayr, 1972). Dietrich and Haider argue that creativity is not Lamarckian because Lamarckian theory does not allow for individuality, waste, or competition:

> A Lamarckian transformational system is… not based on naturally occurring variation but on adaptation-guaranteeing instruction. Importantly, there is no place for individuality or variability in this evolutionary algorithm. Change here acts on every individual unit in the same way. Variation exists, but it is treated as noise and thus evolutionarily unimportant. The Lamarckian system, in other words, includes no waste and no competition.

We know of no writing on Lamarckism (e.g., Burkhardt, 2013; Mayr, 1972) that is consistent with the idea that there is no place for individuality or variability in Lamarckian evolution, that instructions in such a system guarantee adaptation, that change acts on every unit in the same way, or that variation is treated as noise. It is true that non-Darwinian evolutionary processes feature cooperation rather than competition; indeed they are sometimes referred to as "communal exchange" (Vetsigian et al., 2006). However, the authors' claims about the lack of individuality in Lamarckian evolution are at odds with empirical and theoretical studies of microbial life and of the origin and evolution of early life (Gabora, 2006; Kauffman, 1993; Koonin, 2009; Koonin, Makarova, & Aravind, 2001; Koonin & Wolf, 2012; Lynch, 2007; New & Pohorille, 2000; Segre, 2000; Vetsigian, Woese & Goldenfeld, 2006; Wächtershäuser, 1997; Woese, 2002; Woese, Goldenfeld, & 2009; Woese, Olsen, Ibba, & Soll, 2000). It is easy to see why. Consider the period referred to by Vetsigian et al. (2006) as the *Darwinian threshold*, some several hundred million years after the origin of the earliest living protocells, when life shifted from communal exchange to a mode of evolutionary change dominated by natural selection acting largely on lineages. Instead of replicating through haphazard communal interaction amongst autocatalytic protocells and duplication of individual elements followed by budding or cell division, replication started to take place by way of a genetic code largely in lineages. The benefit of shifting to reproduction and selection of heritable variation in a lineage by way of an evolvable genome, exploring DNA, RNA, and protein space, was vastly increased efficiency, but it came at the *cost* of homogenization within asexual lineages; indeed it was not until the advent of sexual reproduction that diversity began to significantly increase.

Another argument Dietrich and Haider make to support the idea that creativity is not Lamarckian is the following:

(1) In a Lamarckian process variation is always adaptive.
(2) Therefore, Lamarckian evolution is totally directed (i.e., 100% coupling of the variation-generating mechanisms to those involved in selection).
(3) Creativity is not totally directed, so it is not Lamarckian.

The claim that a Lamarckian system variation is certain to be adaptive implies that there exists a static, objective evaluation function over a space of possible variants. In fact the merit of a particular variant is in general context-dependent; environmental changes as well as nonlinear interactions amongst components continually alter their contribution to the fitness of the whole in its ecological context (Kauffman, 1993). The claim that Lamarckian evolution is always adaptive is also incompatible with the open-endedness of a space of possibilities; one never knows in advance the possible applications of an idea, for example, so the selective forces that will eventually act upon any idea are unknowable.

The claim that creativity is not Lamarckian because Lamarckian evolution is totally directed is also problematic. A system is Lamarckian not on the basis of *how directed the novelty generating process is* but because there is *transmission of acquired traits across generations*. The 'designer baby' concept is an example of directed evolution, whereas if an organism got wounded and the wound were somehow transmitted to its offspring that would represent an example of Lamarckian evolution.

In short, we know of no theoretical or empirical basis for the authors' claim that creativity cannot be Lamarckian. Therefore, it seems reasonable to consider Lamarckian approaches (e.g., Barry, 2013; Gabora, 2004, 2005, 2013).

**Scaffolding to bypass intermediate forms**

Dietrich and Haider's paper raises the interesting speculation that creativity involves not just *simulation*—which they claim imitates content—but *emulation* involving the scaffolding of thought trials—which additionally involves imitation of the processes by which the content is transformed. They justify their emphasis on the scaffolding of thought trials on the basis that evolution requires, "an instant pay-off requirement", i.e., a variant must be fitter than alternatives or else it is immediately discarded. Scaffolding, they argue, provides a way of meeting this requirement. This is incompatible with standard evolutionary theory, i.e., biologists accept that neutral evolution is widespread and that small populations can stabilize on neutral or even deleterious variants (Kimura, 1983). Indeed, those who adhere to Kimura's neutral theory of evolution believe that genetic drift (change in the frequency of a variant in a population due to random sampling) plays a greater role than selection. Dietrich and Haider's belief that evolution requires an instant pay-off leads them to argue that scaffolding enables cultural evolution trajectories to "bypass impossible intermediates". They write, "An arch is the canonical example of an interlocking design that must leap over non-adaptive, intermediate forms. Biological evolution cannot do that." This incorrectly suggests that biological evolution always gets stuck in local maxima (i.e., cannot move from an intermediate solution to a better solution by way of an inferior solution). In fact, the *strength* of genetic algorithms compared to other search algorithms is that because they exhibit the proverbial "swarm intelligence" of a population-based method they are particularly *adept* at escaping local maxima. Although mental simulation and scaffolding undoubtedly enable creative minds to turn unfinished mediocre ideas into superior ones, evolution does not actually require an instant pay-off. Accordingly, in the evolution of creative ideas (whether it be Darwinian, Lamarckian, or something else), neutral or even deleterious variation can persist for some time.

### Implications of evolutionary theory for dual processing models

As the authors say, a bridging of the prediction literature with research on topics addressed in the creativity literature such as intuition, insight, and the implicit system would be valuable. We question their claim that the "implicit system cannot, by itself, be creative above and beyond the creativity exhibited by evolutionary algorithms in nature – in a blind manner". This may be consistent with how the implicit system is characterized in some dual processing theories but not in others, and particularly those in the creativity literature. Judgments driven by the implicit system can arise due to different types of intuitions, some of which reflect statistical and inferential regularities that are far from blind (Pretz et al., 2014; see Sowden, Pringle, & Gabora, 2014, for a discussion of the relationship between dual processing models and creativity). Here the implicit system is generally characterized as giving rise specifically to ideas that have particular features which, though seemingly "remote" in the given context, are relevant to it (e.g., Finke, Ward, & Smith, 1992). In other words, while the explicit system draws upon patterns of causation, the implicit system draws upon patterns of correlation. This relevance is not a matter of luck or chance but arises as a natural consequence of the architecture of associative memory; ideas that share features are encoded in overlapping distributed sets of cell assemblies, and therefore able to evoke one another through synchronous neural activation even when previously thought to be quite unrelated (Gabora, 2010). Thus we view the claim that the implicit system can only exhibit blind creativity as an area of disagreement.

**A path forward**
Where does this leave us? Though the basic idea of framing creativity in evolutionary terms is promising, the authors' proposal to model creativity as a Darwinian process of trial and error presents theoretical problems, and their suggested implementation using a Bayesian network is not sufficiently fleshed out to enable evaluation. An alternative is to model the evolution of creative ideas as a Lamarckian process, drawing on formal models of concept combination (which lends itself to mathematical modeling more readily than some other creative processes) (e.g., Aerts & Gabora, 2005; Gabora & Aerts, 2009; Gabora, 2013). The approach addresses the challenge of modeling the emergence of new attributes (such as emergence of the concept of a spout during the invention of a teapot) (Gabora, & Carbert, 2015; Gabora, Scott, & Kauffman, 2013). Emergent features are common in creative concept combination (e.g., Estes & Ward, 2002) but to our knowledge they are not currently addressed by the prediction literature nor by any mathematical model of creativity that treats it as variation and selection amongst thought trials (e.g., Simonton, 2013, 2014). Another challenge addressed by the Lamarckian approach is that of modeling the state of a vague, unfinished, or "half-baked" idea. It seems to us that the experience of having an unfinished idea is coherent and unified (despite lack of clarity about how the idea will unfold). To model this experience as a "collection of thought trials" seems problematic because each thought trial is a distinct experience.

**Conclusions**
Dietrich and Haider are on the right track in proposing an integrated foundation for creativity research in which proximate (neuroscientific) explanations are informed by ultimate (evolutionary) mechanisms as well as studies of prediction. Errors such as those pointed out in this commentary are not uncommon in interdisciplinary undertakings such as this one where scholars are likely to tread beyond their primary area of expertise. We believe the gains to be made bridging disciplines more than offset the increased risk of such errors, and it is only by pointing out such errors that a project such as this can move forward. We suggest that shifting from a Darwinian to a Lamarckian framework could be a step toward realizing their vision of grounding creativity in an evolutionary-prediction framework. We applaud the authors for the originality and ambitiousness of this project, which, in broad strokes, points creativity research in a promising and valuable direction.

## Acknowledgements

This research was supported in part by a grant to the first author from the Natural Sciences and Engineering Research Council of Canada. We thank the editor and reviewers for valuable suggestions on the manuscript.